\documentclass[11pt]{article}
\usepackage{moriond,epsfig}
\usepackage{amsfonts}
\usepackage{picinpar}

\begin{document}
\vspace*{4cm}
\title{MUON $g-2$ IN A MODEL WITH ONE EXTRA DIMENSION}

\author{ M.CIRELLI }

\address{Scuola Normale Superiore and INFN,\\ 
Piazza dei Cavalieri 7, Pisa I-56126, Italy}

\maketitle

\abstracts{The computation of the muon anomalous magnetic moment in the framework of a proposed extension of the Standard Model to 5 dimensions is presented. The result (a small correction with respect to the SM prediction) is briefly discussed.}

\vspace{-5mm}
The measurement of the muon anomalous magnetic moment $a_{\mu}$  is currently one of the most stringent tests for ``new physics'' scenarios, particularly in the light of the recent and the future promised results from E821 experiment at BNL~\cite{BNL}.
Models with extra (space) dimensions are among the most interesting of such scenarios, but are often not capable of producing quantitative predictions and, as a consequence, can hardly ever be ruled out or confirmed by present energy experiments.  
In the model proposed in ref.~\cite{BHN}, on the contrary, calculability is achieved for several quantities; in this talk, based on ref.~\cite{g-2}, I present and discuss the computation of $a_{\mu}$.

\section{The framework defined}
\vspace{-3mm}

\begin{window}[0,r,%
{\fbox{\epsfig{file=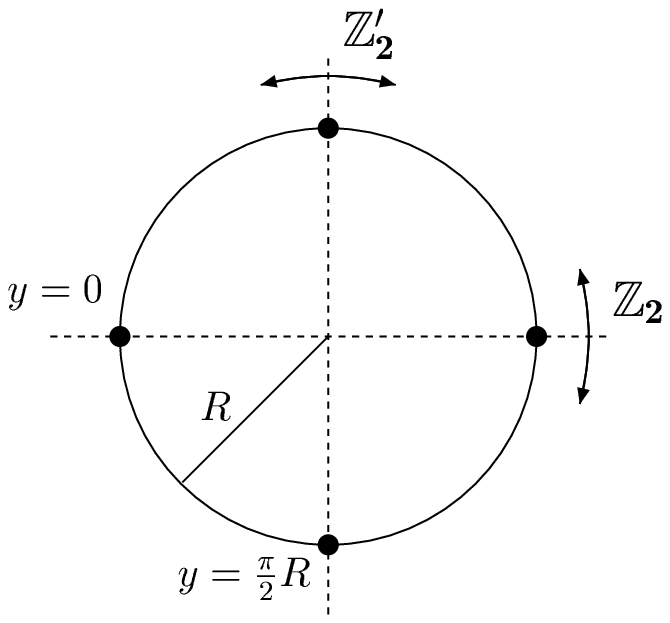, width=55mm}}},]
The theory proposed in ref.~\cite{BHN} is an extension of the Standard Model to 5 dimensions, with $\mathcal{N}=1$ supersymmetry, compactified on $\mathcal{M}^4 \times \mathbb{S}^1/(\mathbb{Z}_{2} \times \mathbb{Z}'_{2})$. The compactification scale $1/R$ is set to $\sim 370 \pm 70$ GeV.~\cite{BHN-FI}

\noindent This means that the gauge group is the SM one and the field content is given by the embedding of the usual SM fields in the extra dimension: for every gauge boson $A^{\mu}$ there is a 5D vector supermultiplet $(A^M,\lambda,\lambda',\sigma)$; for every matter field $Q, U, D, L, E$ and for the single Higgs $H$ there are 5D matter supermultiplets $(\psi,\varphi,\varphi^c,\psi^c)$. 

\noindent As compulsory for a non-abelian gauge theory in 5D, the model posesses a cutoff ${\mathit \Lambda}$, which is set to $\sim 5/R \simeq 1.8$ TeV.  
\end{window}

%\vspace{7mm}

\begin{figwindow}[0,r,%
{\fbox{\epsfig{file=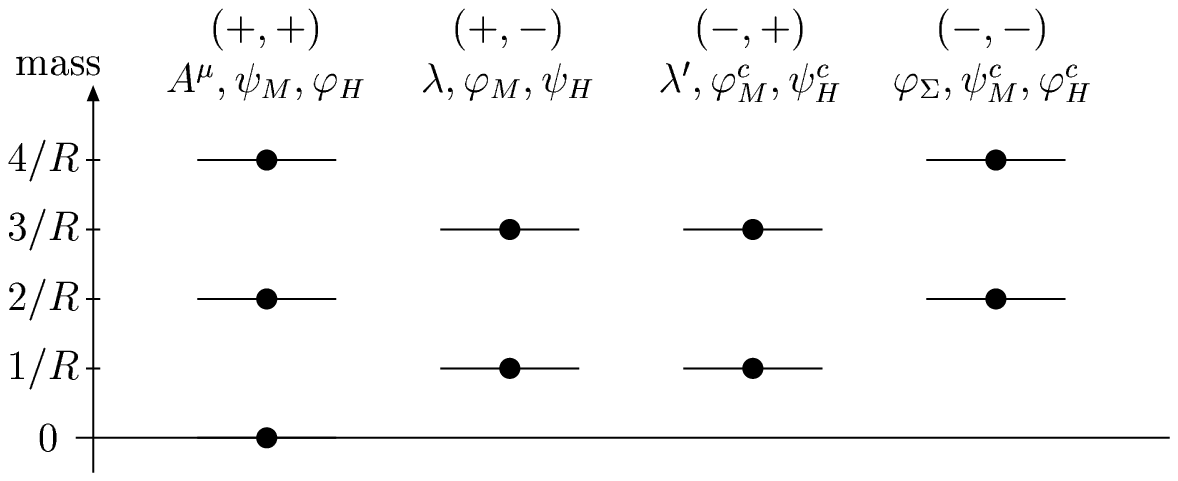, width=95mm}}},Tree-level spectrum of the model\label{fig:spectrum}]
\noindent Under the double orbifolding, pari\-ties are assigned to each field and the resulting spectrum is shown in Fig.\ref{fig:spectrum}: the zero modes reproduce the SM fields but in addition one has to deal with four towers of massive Kaluza-Klein states.

\noindent The global supersymmetry is completely broken, but restricted local supersymmetric transformations still hold. 
\end{figwindow}

\vspace{2mm}

\section{Computation of $g-2$}
\vspace{-3mm}

The muon anomalous magnetic moment $a_{\mu}$ is defined by the effective Lagrangian term

\begin{equation}
\mathcal{L} = \frac{i e}{2 m}\: a_{\mu}\: \big(\overline{\mu} \sigma_{\rho \sigma} F^{\rho \sigma} \mu \big)  \hspace{2cm}  a_{\mu}=\frac{g_{\mu}-2}{2}
\end{equation}

\noindent At one loop, new contributions to $a_{\mu}$ arise from every diagram featuring the muon and a photon as external particles, when the loop is filled with the extra fields of the model by using any allowed vertex (see Figure 1 and Appendix A in ref.~\cite{g-2}).

\noindent We computed all the contributions at first order in $(m_{\mu}R)^2$ and in $(M_WR)^2$, and we checked that higher orders in $(M_WR)^2$ have a negligible impact.
The resummation of the whole KK tower of states is finally performed for any diagram.\footnote{The existence of the cutoff is not inconsistent with such a summation, see refs.~\cite{regularization}.}

\noindent The total correction with respect to the SM prediction for $a_{\mu}$ is found to be

\begin{equation} \label{finalres}
\Delta a_{\mu}^{this~ model} = - \frac{g^2}{192} \frac{m_{\mu}^2}{M_{W}^{2}} \frac{11 -  18~ \sin^2\theta_W}{12~ \cos^2\theta_W} (M_W R)^2 = - (1.1\: _{-0.3}^{+0.6}) \cdot 10^{-10}
\end{equation}

\noindent and is to be compared with the uncertainties of the SM result~\cite{prades} $a_{\mu}^{SM} = (11\: 659\: 179.2 \pm 9.4)\cdot 10^{-10}$.

\vspace{-3mm}
\section{Conclusions}
\vspace{-3mm}

The deviations from the SM value of $a_{\mu}$ are quite small and well inside its errors; in this sense the model under consideration is a {\it viable} extension of the SM.
Moreover, the predictive capability of the model has been shown for this quantity: the computation is {\it reliable}, i.e. insensitive to the cutoff and stable under higher order effects.

\begin{footnotesize}
\section*{Acknowledgments}
\vspace{-2mm}
I thank R.Barbieri and R.Rattazzi for suggestions and discussions that led to the work presented here and I warmly thank G.Cacciapaglia and G.Cristadoro for the fruitful collaboration. It's also a pleasure to thank the Organizing Committee of the Moriond conference.
\end{footnotesize}

\end{document}